\documentstyle[amssymb,prd,aps,epsfig]{revtex}

\newcommand{\be}{\begin{equation}}
\newcommand{\ee}{\end{equation}}
\newcommand{\ba}{\begin{eqnarray}}
\newcommand{\ea}{\end{eqnarray}}

\begin{document}
\draft
\title{{\bf Dimensional Reduction of a Lorentz and CPT-violating
Maxwell-Chern-Simons Model}}
\author{H. Belich Jr.$^{a,b}$, M.M. Ferreira Jr.$^{b,c}$ and J.A. Helay\"{e}l-Neto$%
^{a,b}$, M.T.D. Orlando$^{b,d}$\thanks{{\tt e-mails:} {\tt belich@cbpf.br,
manojr@cbpf.br, helayel@cbpf.br, orlando@cce.ufes.br}}}
\address{$^{a}${\it Centro Brasileiro de Pesquisas F\'{i}sicas (CBPF)},\\
Coordena\c{c}\~{a}o de Teoria de Campos e Part\'{i}culas (CCP), \\
Rua Dr. Xavier Sigaud, 150 - Rio de Janeiro - RJ 22290-180 - Brazil.\\
$^{b}${\it Grupo de F\'{i}sica Te\'{o}rica Jos\'{e} Leite Lopes, }\\
Petr\'{o}polis - RJ - Brazil.\\
$^{c}${\it Universidade Federal do Maranh\~{a}o (UFMA)}, \\
Departamento de F\'{i}sica, Campus Universit\'{a}rio do Bacanga,\\
S\~{a}o Luiz - MA, 65085-580 - Brazil. \\
$^{d}$ {\it Universidade Federal do Esp\'{i}rito Santo (UFES)},\\
Departamento de F\'{i}sica e Qu\'{i}mica, Av. Fernando Ferrarim, S/N\\
Goiabeiras, Vit\'{o}ria - ES, 29060-900 - Brasil}
\maketitle

\begin{abstract}
Taking as starting point a Lorentz and CPT non-invariant Chern-Simons-like
model defined in 1+3 dimensions, we proceed realizing its dimensional
reduction to $D=1+2$. One then obtains a new planar model, composed by the
Maxwell-Chern-Simons (MCS)\ sector, a Klein-Gordon massless scalar field,
and a coupling term that mixes the gauge field to the external vector, $%
v^{\mu }$. In spite of breaking Lorentz invariance in the particle frame,
this model may preserve the CPT symmetry for a single particular choice of $%
v^{\mu }$. Analyzing the dispersion relations, one verifies that the reduced
model exhibits stability, but the causality can be jeopardized by some
modes. The unitarity of the gauge sector is assured without any restriction,
while the scalar sector is unitary only in the space-like case.
\end{abstract}


\section{\ Introduction\-}

In a common sense, it is generally settled that a ``good'' Quantum Field
Theory (QFT) must respect at least two symmetries:\ the Lorentz covariance
and the CPT\ invariance. The traditional framework of a local QFT, from
which one derives the Standard Model that sets the physics inherent to the
fundamental particles, satisfies both these symmetries. In the beginning of
90%
\'{}%
s, a new work \cite{Jackiw} proposing a correction term to the conventional
Maxwell Electrodynamics, that preserves the gauge invariance despite
breaking the Lorentz, CPT and parity symmetries, was first analyzed. The
correction term, composed by the gauge potential, $A_{\mu },$ and an
external background 4-vector, $v_{\mu },$ has a Chern-Simons-like structure, 
$\epsilon ^{\mu \nu \kappa \lambda }v_{\mu }A_{\nu }F_{\kappa \lambda }$,
and is responsible by inducing an optical activity of the vacuum - or
birefringence - among other effects. In this same work, however, it is shown
that astrophysical data do not support the birefringence and impose
stringent limits on the value of the constant vector $v_{\mu },$ reducing it
to a negligible correction term. Similar conclusions, also based on
astrophysical observations, were also confirmed by Goldhaber \& Timble \cite
{Goldhaber}. Some time later, Colladay and Kostelecky \cite{Colladay}
adopted a quantum field theoretical framework to address the issue of CPT-
and Lorentz-breakdown as a spontaneous violation. In this sense, they
constructed an extension to the minimal Standard Model, which maintains
unaffected the $SU(3)\times SU(2)\times U(1)$ gauge structure of the usual
theory, and incorporates the CPT-violation as an active feature of the
effective low-energy broken action. They started from a usual CPT- and
Lorentz-invariant action as defining the properties of what would be an
underlying theory at the Planck scale \cite{Kostelec1}, which then suffers a
spontaneous breaking of both these symmetries. In the broken phase, there
rises the effective action, endowed with breakdown of CPT and Lorentz
symmetries, but conservation of covariance under the perspective of the
observer inertial frame. The Lorentz invariance is spoiled at the level of
the particle-system, which can be viewed in terms of the non-invariance of
the fields under boost and Lorentz rotations (relative inertial
observer-frames). This covariance breakdown is\ also manifest when analyzing
the dispersion relations, extracted from the propagators.

Investigations concerning the unitarity, causality and consistency of a QFT
endowed with violation of Lorentz and CPT symmetries (induced by a
Chern-Simons term) were carried out by Adam \& Klinkhamer \cite{Adam}. As
result, it was verified that the causality and unitarity of this kind of
model can be preserved when the fixed (background) 4-vector is space-like,
and\ spoiled whenever it is time-like or null. A consistency analysis of
this model, carried out in the additional presence of a scalar sector
endowed with spontaneous symmetry breaking (SSB) \cite{Baeta}, has confirmed
the results obtained in ref. \cite{Adam}, that is: the space-like case is
free from unitarity illnesses, which arise in the time- and light-like cases.

The active development of Lorentz- and CPT-violating theories in $D=1+3$ has
come across the inquiry about the structure of a similar model in 1+2
dimensions and its possible implications. In order to study a planar theory,
endowed with Lorentz- and CPT-violation, one has decided to adopt a
dimensional reduction procedure, that is: one starts from the original
Chern-Simons-like term, $\epsilon ^{\mu \nu \kappa \lambda }v_{\mu }A_{\nu
}F_{\kappa \lambda }$, promoting its systematic reduction to $D=1+2,$ which
yields a pure Chern-Simons term and a Lorentz non-invariant mixing term. Our
objective, therefore, is to achieve a planar model, whose structure is
derived from a known counterpart defined in 1+3 dimensions,\ and to
investigate some of its features, like propagators, dispersion relations,
causality, stability and unitarity.

More specifically, one performs the dimensional reduction to 1+2 dimensions
of the Abelian gauge invariant model with non-conservation of the Lorentz
and CPT symmetries \cite{Jackiw}, \cite{Adam} induced by the term $\epsilon
^{\mu \nu \kappa \lambda }v_{\mu }A_{\nu }F_{\kappa \lambda }$, resulting in
a gauge invariant Planar Quantum Electrodynamics (QED$_{3}$) composed by a
Maxwell-Chern-Simons gauge field $\left( A_{\mu }\right) ,$ by a scalar
field $\left( \varphi \right) ,$ a scalar parameter $\left( s\right) $
without dynamics (the Chern-Simons mass), and a fixed 3-vector $\left(
v^{\mu }\right) $. Besides the MCS sector, this Lagrangian has a massless
scalar sector, represented by the field $\varphi $, which also works out as
the coupling constant in the Chern-Simons-like structure that mixes the
gauge field to the 3-vector, $v^{\mu }$ (where one gauge field is replaced
by $v^{\mu }$). This latter term is the responsible by the Lorentz
noninvariance. Therefore, the reduced Lagrangian is endowed with three
coupled sectors: a MCS sector, a massless Klein-Gordon sector and a mixing
Lorentz-violating one. As it is well-known, the MCS sector breaks both
parity and time-reversal \ symmetries, but preserves the Lorentz and CPT\
ones. The scalar sector preserves all discrete symmetries and Lorentz
covariance, whereas the mixing sector, as it will be seen, breaks Lorentz
invariance (in relation to the particle-frame), keeps conserved parity and
charge-conjugation symmetries, but may break or preserve time-reversal
symmetry. This implies that it may occur both conservation (for a purely
space-like $v^{\mu }$) and violation (for $v^{\mu }$ time-like and
light-like) of the CPT\ invariance.

In short, this paper is outlined as follows. In Section II, one accomplishes
the dimensional reduction, that leads to the reduced model. Having
established the new planar Lagrangian, one then devotes some algebraic
effort for the derivation of the propagators of the gauge and scalar fields,
which requires the evaluation of a closed algebra composed by eleven
projector operators, displayed into Table I. In Section III, we investigate
the stability and the causal structure of the theory. One addresses the
causality looking directly at the dispersion relations extracted from the
poles of the propagators, which reveal the existence of both causal and
non-causal modes. All the modes, nevertheless, present positive definite
energy (positivity) relative to any Lorentz frame, which implies an overall
stability. In Section IV, we accomplish the unitarity analysis, based on the
matrix residue evaluated at the poles of the propagators. The unitarity of
the overall model is ensured in the case one adopts a purely space-like
background-vector, $v^{\mu }$. In Section V, we present our Concluding
Comments.

\section{The Dimensionally Reduced Model}

One starts from the Maxwell Lagrangian\footnote{%
Here one has adopted the following metric conventions: $g_{\mu \nu
}=(+,-,-,-)$ in $D=1+3,$ and $g_{\mu \nu }=(+,-,-)$ in $D=1+2$. The greek
letters (with hat) $\hat{\mu},$ run from 0 to 3, while the pure greek
letters, $\mu ,$ run from 0 to 2.} in 1+3 dimensions supplemented by a term
that couples the dual electromagnetic tensor to a fixed 4-vector, $v^{\mu },$
as it appears in ref. \cite{Jackiw}: 
\begin{equation}
{\cal L}_{1+3}=\biggl\{-\frac{1}{4}F_{\hat{\mu}\hat{\nu}}F^{\hat{\mu}\hat{\nu%
}}+\frac{1}{2}\epsilon ^{\hat{\mu}\hat{\nu}\hat{\kappa}\hat{\lambda}}v_{\hat{%
\mu}}A_{\hat{\nu}}F_{\hat{\kappa}\hat{\lambda}}+A_{\hat{\nu}}J^{\hat{\nu}}%
\biggr\},  \label{action1}
\end{equation}
with the additional presence of the coupling between the gauge field and the
external current, $A_{\hat{\nu}}J^{\hat{\nu}}.$ This model (in its free
version) is gauge invariant but does not preserve Lorentz and CPT symmetries
relative to the particle frame. For the observer system, the
Chern-Simons-like term transforms covariantly, once the background also is
changed under an observer boost: $v^{_{\hat{\mu}}}\longrightarrow v^{\hat{\mu%
}^{^{\prime }}}=\Lambda _{\text{ \ }\alpha }^{\mu }v^{\alpha }$. In
connection with the particle-system, however, when one applies a boost on
the particle, the background 4-vector is supposed to remain unaffected,
behaving like a set of four independent numbers, which configures the
breaking of the covariance. This term also breaks the parity symmetry, but
maintain invariance under charge conjugation and time reversal. To study
this model in 1+2 dimensions, one performs its dimensional reduction, \
which consists effectively in adopting the following ansatz over any
4-vector: (i) one keeps unaffected the temporal and also the first two
spatial components; (ii) one freezes the third spacial dimension by
splitting it from the body of the new 3-vector and requiring that the new
quantities $\left( \chi \right) $, defined in 1+2 dimensions, do not depend
on the third spacial dimension:\ $\partial _{_{3}}\chi \longrightarrow 0.$
Applying this prescription\ to the gauge 4-vector, $A^{\hat{\mu}},$ and to
the fixed external 4-vector, $v^{\hat{\mu}},$ and to the 4-current, $J^{\hat{%
\mu}}$, one has: 
\begin{eqnarray}
A^{\hat{\mu}} &\longrightarrow &(A^{\mu };\text{ }\varphi ), \\
v^{\hat{\mu}} &\longrightarrow &(v^{\mu };\text{ }s), \\
J^{\hat{\mu}} &\longrightarrow &(J^{\mu };\text{ }J),
\end{eqnarray}
where: $A^{\left( 3\right) }=\varphi ,$ $v^{\left( 3\right) }=s,$ $J^{(3)}=J$
and $\mu =0,1,2$. According to this process, there appear two scalars: the
scalar field, $\varphi ,$ that exhibits dynamics, and $s,$ a constant scalar
(without dynamics). Carrying out this prescription for eq. (\ref{action1}),
one then obtains: \ \ \ \ \ \ 
\begin{equation}
{\cal L}_{1+2}=-\frac{1}{4}F_{\mu \nu }F^{\mu \nu }+\frac{1}{2}\partial
_{\mu }\varphi \partial ^{\mu }\varphi -\frac{s}{2}\epsilon _{\mu \nu
k}A^{\mu }\partial ^{\nu }A^{k}+\varphi \epsilon _{\mu \nu k}v^{\mu
}\partial ^{\nu }A^{k}-\frac{1}{2\alpha }\left( \partial _{\mu }A^{\mu
}\right) ^{2}+A_{\mu }J^{\mu }+\varphi J,  \label{Lagrange2}
\end{equation}
where the last free term represents the gauge-fixing term, added up after
the dimensional reduction. The scalar field, $\varphi ,$ exhibits a typical
Klein-Gordon massless dynamics and it also appears as the coupling constant
that links the fixed $v^{\mu }$ to the gauge sector of the model, by means
of the new term:\ $\varphi \epsilon _{\mu \nu k}v^{\mu }\partial ^{\nu
}A^{k}.$ In spite of being covariant in form, this kind of term breaks the
Lorentz symmetry in the particle-frame, since the 3-vector $v^{\mu }$ is not
sensitive to particle Lorentz boost, behaving like a set of three scalars.

The Lagrangian (\ref{action1}), originally proposed by Carroll-Field-Jackiw 
\cite{Jackiw}{\it ,} has the property of breaking parity symmetry, even
though conserving time reversal and charge conjugation symmetries, resulting
in nonconservation of the CPT\ symmetry. Simultaneously, the Lorentz
invariance is spoiled, since the fixed 4-vector $v^{\mu }$ breaks the
rotational and boost invariances. On the other hand, the reduced model,
given by eq.(\ref{Lagrange2}), does not necessarily jeopardize the CPT
conservation, which depends truly on the character of the fixed vector $%
v^{\mu }$. As it is known, the parity transformation $\left( {\cal P}\right) 
$ in 1+2 dimensions is characterized by the inversion of only on of the
spatial axis: $x^{\mu }\stackrel{{\cal P}}{\longrightarrow }x^{^{\prime }\mu
}=(x_{o},-x,y),$ the same being valid for the 3-potential: $A^{\mu }%
\stackrel{{\cal P}}{\longrightarrow }A^{^{\prime }\mu }=(A_{0},-A^{\left(
1\right) },A^{\left( 2\right) }).$ The time-reversal transformation $\left( 
{\cal T}\right) $ must keep unchanged the dynamics of the system, so that
one must have: $x^{\mu }\stackrel{{\cal \tau }}{\longrightarrow }x^{^{\prime
}\mu }=(-x_{o},x,y),$ $A^{\mu }\stackrel{{\cal \tau }}{\longrightarrow }%
A^{^{\prime }\mu }=(A_{0},-A^{\left( 1\right) },-A^{\left( 2\right) })$,
while the charge conjugation determines: $x^{\mu }\stackrel{{\cal C}}{%
\longrightarrow }x^{^{\prime }\mu }=x^{\mu },$ $A^{\mu }\stackrel{{\cal C}}{%
\longrightarrow }A^{^{\prime }\mu }=-A^{\mu }.$ One knows that the
Chern-Simons term breaks both parity and time-reversal symmetries and keeps
conserved the charge conjugation, which assures the global CPT\ invariance.
The\ new term, $\varphi \varepsilon _{\mu \nu k}v^{\mu }\partial ^{\nu
}A^{k},$ however, will manifest a non-symmetric behaviour before ${\cal T}$%
-transformation: there will occur conservation if one works with a purely
space-like external vector $\left( v^{\mu }=(0,\overrightarrow{v})\text{ }%
\right) $, or breakdown, if $v^{\mu }$ is purely time-like. Under parity and
charge conjugation transformations, in turn, this term will evidence
non-invariance for any adopted $v^{\mu }$, thereby one can state that it
will occur CPT\ conservation when $v^{\mu }$ is purely space-like, and CPT\
violation otherwise. Here, the field $\varphi $ was considered as having a
scalar character under the parity transformation. Yet, if this field behaves
like a pseudo-scalar\footnote{%
The adoption of a pseudo-scalar field can be justified by looking at the
vector character of the potential ($\overrightarrow{A}\stackrel{{\cal P}}{%
\longrightarrow }-\overrightarrow{A}$) before the dimensional reduction. If
one assumes that the field $\varphi $ maintains the same behaviour of its
ancestral $\left( A_{3}\right) $, one has a pseudo-scalar.}, the CPT
conversation will be assured for a purely time-like $v^{\mu }$. For a
light-like $v^{\mu },$ there will always occur time-reversal non-invariance,
and consequently, CPT violation.

Neglecting divergence terms, one can write the linearized free action in an
explicitly quadratic form, namely: 
\begin{equation}
\Sigma _{1+2}=\int d^{3}x\frac{1}{2}\biggl\{A^{\mu }[M_{\mu \nu }]A^{\nu
}-\varphi \square \varphi +\varphi \left[ \epsilon _{\mu \alpha \nu }v^{\mu
}\partial ^{\alpha }\right] A^{\nu }+A^{\mu }\left[ \epsilon _{\nu \alpha
\mu }v^{\nu }\partial ^{\alpha }\right] \varphi \biggr\},  \label{action3}
\end{equation}
which can also appear in the matrix form:\ 

\begin{equation}
\Sigma _{1+2}=\int d^{3}x\frac{1}{2}\left( 
\begin{array}{cc}
A^{\mu } & \varphi
\end{array}
\right) \left[ 
\begin{array}{cc}
M_{\mu \nu } & T_{\mu } \\ 
-T_{\nu } & -\square
\end{array}
\right] \ \left( 
\begin{array}{c}
A^{\nu } \\ 
\varphi
\end{array}
\right) .  \label{actionQ}
\end{equation}
The action (\ref{actionQ}) has as nucleus a square matrix, $P,$ composed by
the quadratic operators of the initial action. The mass dimension of the
physical parameters and tensors are:\ $\left[ A^{\mu }\right] =\left[
\varphi \right] =1/2,$ $\left[ v^{\mu }\right] =\left[ s\right] =1,\left[
T_{\mu }\right] =\left[ M_{\mu \nu }\right] =2.$ Here, some definitions are
necessary:\ 
\begin{eqnarray}
M_{\mu \nu } &=&\square \theta _{\mu \nu }+s{\rm \ }S_{\mu \nu }+\frac{%
\square }{\alpha }\omega _{\mu \nu },\text{ \ \ \ }T_{\nu }=S_{\mu \nu
}v^{\mu }, \\
\ S_{\mu \nu } &=&\varepsilon _{\mu \kappa \nu }\partial ^{\kappa },\text{ }%
\theta _{\mu \nu }=\eta _{\mu \nu }-\omega _{\mu \nu },\text{ \ \ }\omega
_{\mu \nu }=\frac{\partial _{\mu }\partial _{\nu }}{\square },
\end{eqnarray}
where $\theta _{\mu \nu }$, $\omega _{\mu \nu },S_{\mu \nu }$ stand
respectively for the transverse, longitudinal and Chern-Simons dimensionless
projectors, while $M_{\mu \nu }$ is the quadratic operator associated to the
MCS sector. \ The inverse of the square matrix $P,$ given at the action (\ref
{actionQ}), yields the propagators of the gauge and the scalar fields, which
are also written in a matrix form, the propagator-matrix $\left( \Delta
\right) $: 
\begin{equation}
\Delta =P^{-1}=\frac{-1}{\left( \square M_{\mu \nu }-T_{\mu }T_{\nu }\right) 
}\left[ 
\begin{array}{cc}
-\square & T_{\nu } \\ 
-T_{\mu } & M_{\mu \nu }
\end{array}
\right] ,
\end{equation}
The propagator of the gauge field, $\Delta _{11}$, and of the scalar field, $%
\Delta _{22},$ are written as:\ 
\begin{eqnarray}
\left( \Delta _{11}\right) ^{\mu \nu } &=&\left[ \square \theta _{\mu \nu }+s%
{\rm \ }S_{\mu \nu }+\frac{\square }{\alpha }\omega _{\mu \nu }-\frac{1}{%
\square }T_{\mu }T_{\nu }\right] ^{-1}, \\
\left( \Delta _{22}\right) &=&-\frac{M_{\mu \nu }}{\square }\left[ \square
\theta _{\mu \nu }+s{\rm \ }S_{\mu \nu }+\frac{\square }{\alpha }\omega
_{\mu \nu }-\frac{1}{\square }T_{\mu }T_{\nu }\right] ^{-1}, \\
\left( \Delta _{12}\right) ^{\mu } &=&-\frac{T_{\nu }}{\square }\left[
\square \theta _{\mu \nu }+s{\rm \ }S_{\mu \nu }+\frac{\square }{\alpha }%
\omega _{\mu \nu }-\frac{1}{\square }T_{\mu }T_{\nu }\right] ^{-1}, \\
\left( \Delta _{21}\right) ^{\nu } &=&\frac{T_{\mu }}{\square }\left[
\square \theta _{\mu \nu }+s{\rm \ }S_{\mu \nu }+\frac{\square }{\alpha }%
\omega _{\mu \nu }-\frac{1}{\square }T_{\mu }T_{\nu }\right] ^{-1},
\end{eqnarray}
while the terms $\Delta _{12},$ $\Delta _{21}$ are related to the mixed
propagators $\langle A_{\mu }\varphi \rangle $, $\langle \varphi A_{\mu
}\rangle $ that indicate a scalar mediator turning into a gauge mediator and
vice-versa. Here, for future purposes, it is useful to present the inverse
of the tensor $M_{\mu \nu },$ that is, the propagator of the pure MCS
Lagrangian: 
\begin{equation}
\left( M_{\mu \nu }\right) ^{-1}=\frac{1}{\square +s^{2}}\theta ^{\nu \mu }-%
\frac{s}{\square (\square +s^{2})}S^{\nu \mu }+\frac{\alpha }{\square }%
\omega ^{\nu \mu },  \label{M_inverse}
\end{equation}

To perform the inversion of the operator above, one needs to define some new
operators, since the ones known so far do not form a closed algebra, as it
is shown below: 
\begin{eqnarray}
S_{\mu \nu }T_{\text{ \ }}^{\nu }T^{\alpha } &=&\square v_{\mu }T^{\alpha
}-\lambda T^{\alpha }\partial _{\mu }=\square Q_{\mu }^{\text{ \ }\alpha
}-\lambda \Phi _{\text{ \ }\mu }^{\alpha }, \\
Q_{\mu \nu }Q^{\alpha \nu } &=&T^{2}v^{\alpha }v_{\mu }=T^{2}\Lambda _{\text{
\ }\mu }^{\alpha }, \\
Q_{\mu \nu }\Phi ^{\nu \alpha } &=&T^{2}v_{\mu }\partial ^{\alpha
}=T^{2}\Sigma _{\mu }^{\text{ \ }\alpha },
\end{eqnarray}
where the new operators are: 
\begin{equation}
Q_{\mu \nu }=v_{\mu }T_{\nu },\text{ \ }\Lambda _{\mu \nu }=v_{\mu }v_{\nu },%
\text{ \ \ }\Sigma _{\mu \nu }=v_{\mu }\partial _{\nu },\text{ \ }\Phi _{\mu
\nu }=T_{\mu }\partial _{\nu },
\end{equation}
and, 
\begin{equation}
\lambda \equiv \Sigma _{\mu }^{\;\mu }=v_{\mu }\partial ^{\mu }\;,\;\text{\ }%
T^{2}=T_{\alpha }T^{\alpha }=(v^{2}\square -\lambda ^{2}).
\end{equation}
Their mass dimension are: $\left[ \Lambda _{\mu \nu }\right] =2,$ $\left[
Q_{\mu \nu }\right] =3,$ $\left[ \Sigma _{\mu \nu }\right] =2\;,\;\left[
\Phi _{\mu \nu }\right] =3$.

Three of these new terms exhibit a non-symmetric structure, which leads to
their consideration in pairs, namely: $Q_{\mu \nu },Q_{\nu \mu };$ $\Sigma
_{\mu \nu },\Sigma _{\nu \mu };$ $\Phi _{\mu \nu },\Phi _{\nu \mu }.$ The
inversion of the operator $\Delta _{11}$ will be realized following the
traditional prescription, $\left( \Delta _{11}^{-1}\right) _{\mu \nu }\left(
\Delta _{11}\right) ^{\nu \alpha }=\delta _{\mu }^{\alpha },$ where the
operator $\left( \Delta _{11}\right) ^{\nu \alpha }$ is composed by all the
possible tensor combinations (of rank two) involving $T_{\mu },v_{\mu
},\partial _{\alpha }.$ \ In such way, the proposed propagator will consist,
at a first glance, of eleven terms: 
\begin{equation}
\left( \Delta _{11}\right) ^{\nu \alpha }=a_{1}\theta ^{\nu \alpha
}+a_{2}\omega ^{\nu \alpha }+a_{3}S^{\nu \alpha }+a_{4}\Lambda ^{\nu \alpha
}+a_{5}T^{\nu }T^{\alpha }+a_{6}Q^{\nu \alpha }+a_{7}Q^{\alpha \nu
}+a_{8}\Sigma ^{\nu \alpha }+a_{9}\Sigma ^{\alpha \nu }+a_{10}\Phi ^{\nu
\alpha }+a_{11}\Phi ^{\alpha \nu },
\end{equation}
which are displayed in Table I, where one observes explicitly the closure of
the operator algebra.

\begin{center}
$%
\mathrel{\mathop{%
\begin{tabular}{|c|c|c|c|c|c|c|c|c|c|c|c|}
\hline
 & $\theta _{\mu \nu }$ & $\omega _{\mu \nu }$ & $S_{\mu \nu }$ & $\;\Lambda _{\mu \nu }$ & $T_{\mu }T_{\nu }$ & $Q_{\mu \nu }$ & $Q_{\nu \mu }$ & $\Sigma _{\mu \nu }$ & $\Sigma _{\nu \mu }$ & $\Phi _{\mu \nu }$ & $\Phi _{\nu \mu }$ \\ \hline
$\theta ^{\nu \alpha }$ & $\theta _{\mu }^{\text{ \ }\alpha }$ & $0$ & $S_{\mu }$ $^{\alpha }$ & $%
\begin{tabular}{l}
$\Lambda _{\mu }^{{\rm \ \ }\alpha }+$ \\ 
$-\frac{\lambda }{\square }\Sigma _{\mu }{}^{\alpha }$%
\end{tabular}
$ & $T_{\mu }T^{\alpha }$ & $Q_{\mu }^{\text{ \ }\alpha }$ & $%
\begin{tabular}{l}
$Q_{\text{ }\mu }^{\alpha }+$ \\ 
$-\frac{\lambda }{\square }\Phi _{\mu }^{{\rm \ \ }\alpha }$%
\end{tabular}
$ & $0$ & $%
\begin{tabular}{l}
$\Sigma _{\text{ \ }\mu }^{\alpha }+$ \\ 
$-\lambda \square \omega _{\text{ \ }\mu }^{\alpha }$%
\end{tabular}
$ & $0$ & $\Phi _{\text{ \ }\mu }^{\alpha }$ \\ \hline
$\omega ^{\nu \alpha }$ & $0$ & $\omega _{\mu }^{\text{ \ }\alpha }$ & $0$ & $\frac{\lambda }{\square }\Sigma _{\mu }{}^{\alpha }$ & $0$ & $0$ & $\frac{\lambda }{\square }\Phi _{\mu }^{{\rm \ \ }\alpha }$ & $\Sigma _{\mu }^{\text{ \ }\alpha }$ & $\lambda \omega _{\text{ \ }\mu }^{\alpha }$ & $\Phi _{\text{ \ }\mu }^{\alpha }$ & $0$ \\ \hline
$S^{\nu \alpha }$ & $S_{\mu }^{\text{ \ }\alpha }$ & $0$ & $-\square \theta _{\mu }^{\text{ \ }\alpha }$ & $Q_{\mu }$ $^{\alpha }$ & $%
\begin{tabular}{l}
$\lambda \Phi _{\mu }^{\text{ \ }\alpha }+$ \\ 
$-\square Q_{\text{ }\mu }^{\alpha }$%
\end{tabular}
$ & $%
\begin{tabular}{l}
$\lambda \Sigma _{\mu }^{\text{ \ }\alpha }+$ \\ 
$-\Lambda _{\mu }^{{\rm \ \ }\alpha }\square $%
\end{tabular}
$ & $-T_{\mu }T^{\alpha }$ & $0$ & $\partial _{\mu }T^{\alpha }$ & $0$ & $%
\begin{tabular}{l}
$\square (\omega _{\mu }$ $^{\alpha }+$ \\ 
$-$ $\Sigma ^{\alpha }$ $_{\mu })$%
\end{tabular}
$ \\ \hline
$\;\Lambda ^{\nu \alpha }$ & $%
\begin{tabular}{l}
$\Lambda _{\mu }^{{\rm \ \ }\alpha }+$ \\ 
$-\frac{\lambda }{\square }\Sigma {}_{\text{ \ }\mu }^{\alpha }$%
\end{tabular}
$ & $\frac{\lambda }{\square }\Sigma {}_{\text{ \ }\mu }^{\alpha }$ & $-Q_{\text{ \ }\mu }^{\alpha }$ & $v^{2}\Lambda _{\mu }^{\text{ \ }\alpha }$ & $0$ & $0$ & $v^{2}Q_{\text{ \ }\mu }^{\alpha }$ & $\lambda \Lambda _{\mu }^{\text{ \ }\alpha }$ & $v^{2}\Sigma _{\text{ \ }\mu }^{\alpha }$ & $\lambda Q_{\text{ \ }\mu }^{\alpha }$ & $0$ \\ \hline
$T^{\nu }T^{\alpha }$ & $T_{\mu }T^{\alpha }$ & $0$ & $%
\begin{tabular}{l}
$\square Q_{\mu }^{\text{ \ }\alpha }+$ \\ 
$-\lambda \Phi _{\text{ \ }\mu }^{\alpha }$%
\end{tabular}
$ & $0$ & $T^{2}T_{\mu }T^{\alpha }$ & $T^{2}Q_{\mu }^{\alpha }$ & $0$ & $0$ & $0$ & $0$ & $T^{2}Q_{\mu }^{\text{ \ }\alpha }$ \\ \hline
$Q^{\nu \alpha }$ & $%
\begin{tabular}{l}
$Q_{\mu }^{\text{ \ }\alpha }+$ \\ 
$-\frac{\lambda }{\square }\Phi _{\text{ \ }\mu }^{\alpha }$%
\end{tabular}
$ & $\frac{\lambda }{\square }\Phi _{\text{ \ }\mu }^{\alpha }$ & $-T_{\mu }T^{\alpha }$ & $v^{2}Q_{\mu }^{\text{ \ }\alpha }$ & $0$ & $0$ & $v^{2}T_{\mu }T^{\alpha }$ & $\lambda Q_{\mu }^{\text{ \ }\alpha }$ & $v^{2}\partial _{\mu }T^{\alpha }$ & $\lambda T_{\mu }T^{\alpha }$ & $0$ \\ \hline
$Q^{\alpha \nu }$ & $Q_{\text{ \ }\mu }^{\alpha }$ & $0$ & $%
\begin{tabular}{l}
$\square \Lambda _{\mu }^{\text{ \ }\alpha }+$ \\ 
$-\lambda \Sigma _{\text{ \ }\mu }^{\alpha }$%
\end{tabular}
$ & $0$ & $T^{2}Q_{\text{ \ }\mu }^{\alpha }$ & $T^{2}\Lambda _{\text{ \ }\mu }^{\alpha }$ & $0$ & $0$ & $0$ & $0$ & $T^{2}\Sigma _{\text{ \ }\mu }^{\alpha }$ \\ \hline
$\Sigma ^{\nu \alpha }$ & $%
\begin{tabular}{l}
$\Sigma _{\mu }{}^{\alpha }+$ \\ 
$-\lambda \omega _{\mu }{}^{\alpha }$%
\end{tabular}
$ & $\lambda \omega _{\mu }^{\text{ \ }\alpha }$ & $-\Phi _{\mu }^{\text{ \ }\alpha }$ & $v^{2}\Sigma _{\mu }^{\text{ \ }\alpha }$ & $0$ & $0$ & $v^{2}\Phi _{\mu }^{\text{ \ }\alpha }$ & $\lambda \Sigma _{\mu }{}^{\alpha }$ & $v^{2}\Lambda _{\mu }^{\text{ \ }\alpha }$ & $\lambda \Phi _{\mu }^{\text{ \ }\alpha }$ & $0$ \\ \hline
$\Sigma ^{\alpha \nu }$ & $0$ & $\Sigma _{\text{ \ }\mu }^{\alpha }$ & $0$ & $\lambda \Lambda _{\mu }^{\text{ \ }\alpha }$ & $0$ & $0$ & $\lambda Q_{\text{ \ }\mu }^{\alpha }$ & $\square \Lambda _{\mu }$ $^{\alpha }$ & $v^{2}\Lambda _{\mu }^{\text{ \ }\alpha }$ & $\square Q_{\mu }^{\text{ \ }\alpha }$ & $0$ \\ \hline
$\Phi ^{\nu \alpha }$ & $\Phi _{\mu }^{\text{ \ }\alpha }$ & $0$ & $%
\begin{tabular}{l}
$\square (\Sigma _{\mu }^{\text{ \ }\alpha }+$ \\ 
$-\lambda \omega _{\mu }{}^{\alpha })$%
\end{tabular}
$ & $0$ & $T^{2}\Phi _{\mu }^{\text{ \ }\alpha }$ & $T^{2}\Sigma _{\mu }{}^{\alpha }$ & $0$ & $0$ & $0$ & $0$ & $\square T^{2}\omega _{\mu }^{\alpha }$ \\ \hline
$\Phi ^{\alpha \nu }$ & $0$ & $\Phi _{\text{ \ }\mu }^{\alpha }$ & $0$ & $\lambda \Phi _{\text{ \ }\mu }^{\alpha }$ & $0$ & $0$ & $\lambda T_{\mu }T^{\alpha }$ & $\square \Phi ^{\alpha }$ $_{\mu }$ & $\lambda \Phi _{\text{ \ }\mu }^{\alpha }$ & $\square T_{\mu }T^{\alpha }$ & $0$ \\ \hline
\end{tabular}
}\limits_{\text{Table I: Multiplicative operator algebra fulfilled by }\theta \text{, }\omega \text{, \thinspace }S\text{,\thinspace\ }\Lambda \text{,}TT\text{, }Q\text{, }\Sigma \text{,and }\Phi \text{. The products are supposed to be in the ordering ``row times column''.}}}%
$\vspace{2mm}
\end{center}

\bigskip Using the data contained in Table I, one finds out that the
gauge-field propagator assumes the form:

\begin{eqnarray*}
\left( \Delta _{11}\right) ^{\mu \nu } &=&\frac{1}{\square +s^{2}}\theta
^{\mu \nu }+\frac{\alpha (\square +s^{2})\boxtimes -\lambda ^{2}s^{2}}{%
\square (\square +s^{2})\boxtimes }\omega ^{\mu \nu }-\frac{s}{\square
(\square +s^{2})}S^{\mu \nu }-\frac{s^{2}}{(\square +s^{2})\boxtimes }%
\Lambda ^{\mu \nu }+\frac{1}{(\square +s^{2})\boxtimes }T^{\mu }T^{\nu } \\
&&-\frac{s}{(\square +s^{2})\boxtimes }Q^{\mu \nu }+\frac{s}{(\square
+s^{2})\boxtimes }Q^{\nu \mu }+\frac{\lambda s^{2}}{\square (\square
+s^{2})\boxtimes }\Sigma ^{\mu \nu }+\frac{\lambda s^{2}}{\square (\square
+s^{2})\boxtimes }\Sigma ^{\nu \mu }-\frac{s\lambda }{\square (\square
+s^{2})\boxtimes }\Phi ^{\mu \nu } \\
&&+\frac{s\lambda }{\square (\square +s^{2})\boxtimes }\Phi ^{\nu \mu },
\end{eqnarray*}
where: $\boxtimes =(\square ^{2}+s^{2}\square -T^{2}).$

By the same procedure, one evaluates the mixed propagator, \ $\left( \Delta
_{12}\right) ^{\alpha }=-\frac{T_{\nu }}{\square }\left( \Delta _{11}\right)
^{\nu \alpha }$, which can be written in the following form:\ 

\begin{equation}
\left( \Delta _{12}\right) ^{\nu }=-\frac{1}{\boxtimes }\left[ T^{\nu
}+sv^{\nu }-\frac{s\lambda }{\square }\partial ^{\nu }\right] ,
\end{equation}
whereas the propagator $\left( \Delta _{21}\right) ^{\nu }$, in turn,
results equal to: 
\[
\left( \Delta _{21}\right) ^{\nu }=-\frac{1}{\boxtimes }\left[ -T^{\nu
}+sv^{\nu }-\frac{s\lambda }{\square }\partial ^{\nu }\right] , 
\]
In order to compute the propagator of the scalar field, 
\begin{equation}
\left( \Delta _{22}\right) =-\frac{1}{\square }\left[ 1-\frac{1}{\square }%
T_{\mu }\left( M_{\mu \nu }\right) ^{-1}T_{\nu }\right] ^{-1},
\end{equation}
one makes use of the inverse of the tensor $M_{\mu \nu }$, given by eq. (\ref
{M_inverse}), so that: $T_{\mu }(M^{-1})^{\mu \nu }T_{\nu }=\left( \square
+s^{2}\right) ^{-1}T^{2}.$ In such a way, a compact scalar propagator arises:

\begin{equation}
\left( \Delta _{22}\right) \text{ }=-\frac{\square +s^{2}}{\boxtimes }
\end{equation}
In momentum-space, the photon propagator takes the final expression: 
\begin{eqnarray}
\text{ }\langle A^{\mu }\left( k\right) A^{\nu }\left( k\right) \rangle 
\text{ } &=&i\biggl\{-\frac{1}{k^{2}-s^{2}}\theta ^{\mu \nu }-\frac{\alpha
(k^{2}-s^{2})\boxtimes (k)+s^{2}\left( v.k\right) ^{2}}{k^{2}(k^{2}-s^{2})%
\boxtimes (k)}\omega ^{\mu \nu }-\frac{s}{k^{2}(k^{2}-s^{2})}S^{\mu \nu }+%
\frac{s^{2}}{(k^{2}-s^{2})\boxtimes (k)}\Lambda ^{\mu \nu }  \nonumber \\
&&-\frac{1}{(k^{2}-s^{2})\boxtimes (k)}T^{\mu }T^{\nu }+\frac{s}{%
(k^{2}-s^{2})\boxtimes (k)}Q^{\mu \nu }-\frac{s}{(k^{2}-s^{2})\boxtimes (k)}%
Q^{\nu \mu }+\frac{is^{2}\left( v.k\right) }{k^{2}(k^{2}-s^{2})\boxtimes (k)}%
\Sigma ^{\mu \nu }  \nonumber \\
&&+\frac{is^{2}\left( v.k\right) }{k^{2}(k^{2}-s^{2})\boxtimes (k)}\Sigma
^{\nu \mu }-\frac{is\left( v.k\right) }{k^{2}(k^{2}-s^{2})\boxtimes (k)}\Phi
^{\mu \nu }+\frac{is\left( v.k\right) }{k^{2}(k^{2}-s^{2})\boxtimes (k)}\Phi
^{\nu \mu }\biggr\},  \label{Prop_A}
\end{eqnarray}
while the scalar and the mixed propagators read as: 
\begin{equation}
\text{ }\langle \varphi \varphi \rangle \text{ }=\frac{i}{\boxtimes (k)}%
\left[ k^{2}-s^{2}\right] ,  \label{Prop_phi}
\end{equation}
\begin{equation}
\langle A^{\alpha }\left( k\right) \varphi \rangle =-\frac{i}{\boxtimes (k)}%
\left[ T^{\alpha }+sv^{\alpha }-\frac{s\left( v.k\right) }{k^{2}}k^{\alpha }%
\right] ,  \label{Prop_Aphi}
\end{equation}

\begin{equation}
\langle \varphi A^{\alpha }\left( k\right) \rangle =-\frac{i}{\boxtimes (k)}%
\left[ -T^{\alpha }+sv^{\alpha }-\frac{s\left( v.k\right) }{k^{2}}k^{\alpha }%
\right] ,  \label{Prop_phiA}
\end{equation}
where: $\boxtimes (k)=\left[ k^{4}-\left( s^{2}-v.v\right) k^{2}-\left(
v.k\right) ^{2}\right] .$ By the above expressions, one notes that the
factor $\boxtimes $ is present on the denominator of all propagators, in
such a way the scalar and the\ gauge field will share the pole structure,
and consequently, the physical excitations associated with the poles of $%
\boxtimes (k).$ This common dependence on $1/\boxtimes $ also amounts to
similarities on the causal structure of the scalar and gauge sectors of this
model, as it will de discussed in Section III.

\section{\ Dispersion Relations, Stability and Causality Analysis}

Some references in literature \cite{Adam},\cite{Andrianov},\cite{Kostelec3}
have dealt with the issue of stability, causality and unitarity concerning
to Lorentz- and CPT-violating theories. The causality is usually addressed
as a quantum feature that requires the commutation between observables
separated by a space-like interval, which one calls microcausality in field
theory \cite{Pauli}. In this section, however, one analyzes causality under
a classical tree-level perspective, in which it is related to the positivity
of a usual Lorentz invariant, $k^{2}.$ The starting-point of all
investigation is the propagator, whose poles are associated to dispersion
relations (DR) that provide informations about the stability and causality
of the model. The causality analysis is then related to the sign of the
propagator poles, given in terms of $k^{2},$ in such a way one must have $%
k^{2}\geq 0$ in order to preserve it (circumventing the existence of
tachyons). In the second quantization framework, stability is related to the
energy positivity of the Fock states for any momentum. Here, stability is
directly associated with the energy positivity of each mode read off \ from
the DR.

The field propagators, given by eqs. (\ref{Prop_A}, \ref{Prop_Aphi}, \ref
{Prop_phi}), present three families of poles at $k^{2}$: 
\begin{equation}
k^{2}=0;\text{ \ }k^{2}-s^{2}=0;\text{ \ }k^{4}-(s^{2}-v.v)k^{2}-(v.k)^{2}=0,
\end{equation}
from which one straightforward infers the DR derived from the Lagrangian (%
\ref{Lagrange2}), namely:\ 
\begin{equation}
k_{0\left( 1\right) }^{2}=\overrightarrow{k}^{2};\text{ \ }k_{0\left(
2\right) }^{2}=\overrightarrow{k}^{2}+s^{2};\text{ \ \ }k_{0\left( 3\right)
}^{2}=\overrightarrow{k}^{2}+\frac{1}{2}\left[ \left( s^{2}-v.v\right) \pm 
\sqrt{\left( s^{2}-v.v\right) ^{2}+4\left( v.k\right) ^{2}}\right] .
\end{equation}
The first dispersion relation, $k_{0}=\pm \overrightarrow{|k}|,$ stands for
a massless photon mode, which carries no degree of freedom, since the
Lagrangian (\ref{Lagrange2}) involves a massive photon. The second DR
represents the Chern-Simons massive mode, $k_{0}=\pm \sqrt{s^{2}+|%
\overrightarrow{k}|^{2}},$ which propagates only one degree of freedom (in
the Maxwell-Chern-Simons electrodynamics the scalar magnetic field encloses
all information of the electromagnetic field, which justifies the existence
of a single degree of freedom). These first two poles apparently respect the
causality condition, since $k^{2}\geq 0$ for them. Once the causality is set
up, the stability comes up as a direct consequence.

Concerning the third DR, corresponding to the roots of $\boxtimes (k)$, it
may provide\ both massless and massive modes for some specific $%
\overrightarrow{k}$-values, but in general, the mode is massive. By
remembering that $\overrightarrow{k}$ is the transfer momentum, whose values
are generally integrated from zero to infinity, one concludes it does not
make much sense to fix any value for $\overrightarrow{k}$ in order to obtain
a particular dispersion relation. Remarking that the term $\boxtimes (k)$ is
ubiquitous in the denominator of all propagators, as it is explicit in eqs. (%
\ref{Prop_A}),(\ref{Prop_phi}), (\ref{Prop_Aphi}), one concludes the causal
structure entailed to the poles of $1/\boxtimes $ will be common to these
three propagators. Specifically, for a purely space-like 3-vector, $v^{\mu
}=(0,\overrightarrow{v}),$ this DR is written as, 
\begin{equation}
\ k_{0\pm }^{2}=\overrightarrow{k}^{2}+\frac{1}{2}\left[ \left( s^{2}+%
\overrightarrow{v}^{2}\right) \pm \sqrt{\left( s^{2}+\overrightarrow{v}%
^{2}\right) ^{2}+4\left( \overrightarrow{v}.\overrightarrow{k}\right) ^{2}}%
\right] .  \label{DR2}
\end{equation}
A simple analysis of this expression indicates that both $\ k_{0+}^{2}$ and $%
k_{0-}^{2}$ are positive-energy modes for any $\overrightarrow{k}$-value
(and for any Lorentz observer), which assures the stability of these modes.
This fact may suggest that the causal structure of the space-like sector of
this model remains preserved, as it was observed by Adam \& Klinkhamer \cite
{Adam} in the context of the 4-dimensional version of this theory, that is
endowed with a dispersion relation very similar to eq. (\ref{DR2}) (this
conclusion was also supported by the attainment of a group velocity,
associated to this mode, smaller than 1). Concerning the pole analysis,
although, we have $k_{+}^{2}>0$ for arbitrary $\overrightarrow{k}$ and $%
k_{-}^{2}<0$ (unless $\overrightarrow{k}\bot \overrightarrow{v}$ or $%
\overrightarrow{k}=0$, which implies $k_{-}^{2}=0$). So, while the mode $%
k_{+}^{2}$ preserves the causality and stability, the mode $k_{-}^{2}$, in
spite of assuring stability, will be in general non-causal, preserving
causality only when $\overrightarrow{k}\bot \overrightarrow{v}$ or $%
\overrightarrow{k}=0.$

In the case of a purely time-like 3-vector, $v^{\mu }=(v_{0},\overrightarrow{%
0}),$ the DR assumes the form:\ 
\begin{equation}
k_{0\pm }^{2}=\frac{1}{2}\left[ \left( s^{2}+2\overrightarrow{k}^{2}\right)
\pm \sqrt{s^{4}+4v_{o}^{2}\overrightarrow{k}^{2}}\right] ,  \label{DR3}
\end{equation}
where one observes a similar behaviour: the mode $k_{0+}^{2}$ will exhibit
stability and causality, while the mode $k_{0-}^{2}$ will present energy
positivity (for arbitrary $\overrightarrow{k}$-value) whenever the
condition, $s^{2}-v_{0}^{2}>0$, is fulfilled. From now on, one must assume
the validity of this condition, so that the mode $k_{0-}^{2}$ can be taken
stable. This latter mode\ is non-causal for any $\overrightarrow{k}\neq 0.$
Assuming the coefficients for Lorentz violation are small near the
Chern-Simons mass $\left( s^{2}\gg v_{0}^{2},|\overrightarrow{v}|^{2}\right)
,$ we obtain an entirely causal theory (at least at zero order in $%
v^{2}/s^{2}).$ This is consistent with some results \cite{Kostelec3}
concerning some quantum theories containing Lorentz-violating terms, which
evidence the preservation of causality when the breaking factors are small.

Hence, the modes $k_{0\pm }^{2}$ exhibit positive energy both in space- and
time-like cases, which also implies these two modes can be written as an
expansion in terms of positive and negative frequency terms. This separation
allows the definition of particles and antiparticles states, a necessary
condition for the quantization of this theory. Nevertheless, the existence
of non-causal modes, both in time- and space-like case, may be seen already
at classical level, as a prediction on the impossibility to realize a
consistent quantization of this model, an issue that will be properly
addressed when one analyses the unitarity at these non-causal poles.
Therefore, the existence of quantization illness will be solved by
investigating the unitarity of the model, matter to be discussed in the next
section.

In a Lorentz covariant framework, $k^{2}$ is a Lorentz scalar, which assures
a unique value for all Lorentz frames. In such a way, if $k^{2}$ represents
a causal mode for one observer, so it will be for all ones. The fact that $%
k^{2}$ has not a positive definite value in an arbitrary Lorentz frame is a
unequivocal indicative of the Lorentz covariance breakdown.

\section{Unitarity Analysis}

In order to analyze the unitarity of the model at tree-level, one has
adopted the method which consists in saturating the propagators with
external currents. The fact that our model possesses two sectors (the scalar
and gauge one) implies that we must saturate the scalar-propagator and the
gauge-propagator individually. In such a way, we write the two saturated
propagators, namely: 
\begin{eqnarray*}
SP_{\langle A_{\mu }A_{\nu }\rangle } &=&J^{\ast \mu }\langle A_{\mu
}(k)A_{\nu }(k)\rangle \text{ }J^{\nu }, \\
SP_{\langle \varphi \varphi \rangle } &=&J^{\ast }\langle \varphi \varphi
\rangle J,
\end{eqnarray*}
where the gauge current $J^{\mu }$ must obey the conservation law valid for
the gauge sector of the system\footnote{%
By applying the differential operator, $\partial _{\mu },$ on the equation
of motion derived from Lagrangian (\ref{Lagrange2}), there results the
following equation (see ref. \cite{Manojr}) for the gauge current: $\partial
_{\mu }J^{\mu }=-\varepsilon ^{\mu \nu \rho }\partial _{\mu }v_{\nu
}\partial _{\rho }\varphi $, which reduces to the conventional
current-conservation law, $\partial _{\mu }J^{\mu }=0$, whenever $v^{\mu }$
is constant or has a null rotational $(\varepsilon ^{\mu \nu \rho }\partial
_{\mu }v_{\nu }=0)$.}, whereas the scalar current, $J,$ is not subject to
any constraint. The unitarity analysis is based on the residues of $SP,$
precisely: the unitarity is ensured whenever the imaginary part of the
residues of $SP$ at the poles of each propagator is positive. It is easy to
notice that the saturated propagator in the momentum-space is the
current-current transition amplitude.

\subsection{Scalar Sector}

We can initiate our analysis by the scalar sector, whose saturated amplitude
is given by:\ $SP_{\langle \varphi \varphi \rangle }=J^{\ast }\langle
\varphi \varphi \rangle J,$ or more explicitly: 
\[
SP_{\langle \varphi \varphi \rangle }=J^{\ast }\frac{i(k^{2}-s^{2})}{%
\boxtimes (k)}J. 
\]
This expression presents two poles, $k_{+}^{2},k_{-}^{2}$, the roots of $%
\boxtimes (k)=0.$ At the purely time-like case, $v^{\mu }=(v_{0},%
\overrightarrow{0}),$ these poles are exactly the ones given by eq. (\ref
{DR3}): $k_{\pm }^{2}=$ $\frac{1}{2}\left[ s^{2}\pm \sqrt{s^{4}+4v_{o}^{2}%
\overrightarrow{k}^{2}}\right] $. Evaluating the residues of $SP_{\langle
\varphi \varphi \rangle }$ at the pole $k_{+}^{2}$ one achieves a positive
imaginary result, while at the pole $k_{-}^{2}$ a positive result appears
only when the condition $\overrightarrow{k}^{2}<(v_{0}^{2}+s^{2})$. In such
a way, one concludes that the unitarity of the scalar sector, in the
time-like case, is not assured. Considering now the purely space-like case, $%
v^{\mu }=(0,\overrightarrow{v}),$ the poles of $SP_{\langle \varphi \varphi
\rangle }$ are given by eq. (\ref{DR2}): $k_{\pm }^{2}=\frac{1}{2}\left[
\left( s^{2}+\overrightarrow{v}^{2}\right) \pm \sqrt{\left( s^{2}+%
\overrightarrow{v}^{2}\right) ^{2}+4\left( \overrightarrow{v}.%
\overrightarrow{k}\right) ^{2}}\right] $. The residues associated with these
two poles exhibit a positive definite imaginary part, so that one can state
that the unitarity of the scalar sector, at the space-like case, is
generically preserved.

\subsection{Gauge-Field sector}

The continuity equation, $\partial _{\mu }J^{\mu }=0,$ in the k-space is
read as: $k_{\mu }J^{\mu }=0$; it allows us to write the current in the
form: $J^{\mu }=(j^{0},0,\frac{k_{0}}{k_{2}}j^{(0)}).$ The conservation
constraint, $\ j^{(2)}=\frac{k_{0}}{k_{2}}j^{(0)},$ appears whenever one
adopts $k^{\mu }=(k_{0},0,k_{2})$ as the momentum. The current conservation
law also reduces to six the number of terms of the photon propagator that
contributes to the evaluation of the saturated propagator:\ 
\begin{equation}
SP_{\langle A_{\mu }A_{\nu }\rangle }=J_{\mu }^{\ast }(k)\biggl\{\frac{i}{D}%
\left( \square \boxtimes g^{\mu \nu }-s\boxtimes S^{\mu \nu }-s^{2}\square
\Lambda ^{\mu \nu }+\square T^{\mu }T^{\nu }-s\square Q^{\mu \nu }+s\square
Q^{\nu \mu }\right) \biggr\}J_{\nu }(k),
\end{equation}
where: $D=\square (\square +s^{2})\boxtimes .$ Writing this expression in
the momentum-space, one obtains: 
\begin{equation}
SP_{\langle A_{\mu }A_{\nu }\rangle }=J^{\ast \mu }(k)\biggl\{iB_{\mu \nu }%
\biggr\}J^{\nu }(k),
\end{equation}
where: $D=k^{2}(k^{2}-s^{2})\boxtimes ,$ with: $\boxtimes \left( k\right)
=k^{4}-(s^{2}-v.v)k^{2}-\left( v.k\right) ^{2}$.

\subsubsection{Time-like case:}

We start \ by analyzing the unitarity in the case corresponding to a
time-like background-vector: $v^{\mu }=(v_{0},\overrightarrow{0}).$ In this
situation, the 2-rank tensor $B_{\mu \nu }$ can be put in the form: 
\begin{equation}
B_{\mu \nu }(k)=\frac{1}{D(k)}\left[ 
\begin{array}{ccc}
k^{2}(s^{2}v_{0}^{2}-\boxtimes ) & 
\begin{array}{c}
ik^{\left( 2\right) }\left( s\boxtimes -v_{0}^{2}s^{2}k^{2}\right)
\end{array}
& ik^{\left( 1\right) }(-s\boxtimes +v_{0}^{2}s^{2}k^{2}) \\ 
ik^{\left( 2\right) }\left( -s\boxtimes +v_{0}^{2}s^{2}k^{2}\right) & 
\begin{array}{c}
k^{2}\left( \boxtimes +v_{0}^{2}k_{2}^{2}\right)
\end{array}
& is\boxtimes k^{\left( 0\right) }-v_{0}^{2}k^{2}k^{\left( 1\right)
}k^{\left( 2\right) } \\ 
ik^{\left( 1\right) }(s\boxtimes -v_{0}^{2}s^{2}k^{2}) & 
\begin{array}{c}
-is\boxtimes k^{\left( 0\right) }-v_{0}^{2}k^{2}k^{\left( 1\right)
}k^{\left( 2\right) }
\end{array}
& k^{2}\left( \boxtimes +v_{0}^{2}k_{1}^{2}\right)
\end{array}
\right] ,
\end{equation}
where: $\boxtimes =k^{4}-(s^{2}-v_{0}^{2})k^{2}-k_{0}^{2}$.

For the pole $k^{2}=0,$ with $k^{\mu }=(k_{0},0,k_{0})$, we have the\
following residue matrix: 
\begin{equation}
B_{\mu \nu }|_{(k^{2}=0)}=\frac{1}{s^{2}}\left[ 
\begin{array}{ccc}
0 & 
\begin{array}{c}
-isk_{0}
\end{array}
& 0 \\ 
isk_{0} & 0 & -isk_{0} \\ 
0 & 
\begin{array}{c}
isk_{0}
\end{array}
& 0
\end{array}
\right] ,  \label{k2zero}
\end{equation}
which is reduced to a null matrix when saturated with the conserved current, 
$J^{\mu }=(j^{0},0,\frac{k_{0}}{k_{2}}j^{(0)}),$ implying also a null
saturation $(SP=0)$. This fact indicates that the mode associated with the
pole $k^{2}=0$ carries no physical degree of freedom, and further, it does
not jeopardize the unitarity.

For the pole $k^{2}=s^{2},$ with $k^{\mu }=(k_{0},0,k_{2})$, the matrix
takes the form,

\begin{equation}
B_{\mu \nu }|_{(k^{2}=s^{2})}=-\frac{1}{s^{2}k_{2}^{2}}\left[ 
\begin{array}{ccc}
s^{2}k_{0}^{2} & -isk^{\left( 2\right) }k_{0}^{2} & 0 \\ 
isk^{\left( 2\right) }k_{0}^{2} & 0 & -isk_{0}k_{2}^{2} \\ 
0 & isk_{0}k_{2}^{2} & -s^{2}k_{2}^{2}
\end{array}
\right] ,  \label{k2s}
\end{equation}
This matrix, whenever saturated with the external current $J^{\mu }=(j^{0},0,%
\frac{k_{0}}{k_{2}}j^{(0)}),$ leads to a trivial saturation\ $\left(
SP=0\right) $, which is compatible with unitarity requirements. The
vanishing of the current-current amplitude at this pole indicates that the
massive excitation $k^{2}=s^{2}$ is not dynamical for the time-like
background.

At the pole $k_{+}^{2}=$ $\frac{1}{2}\left[ s^{2}+\sqrt{s^{4}+4v_{o}^{2}%
\overrightarrow{k}^{2}}\right] ,$ the residue matrix reads as:

\begin{equation}
B_{\mu \nu }|_{(k^{2}=k_{+}^{2})}=\frac{v_{0}^{2}}{%
(k_{+}^{2}-s^{2})(k_{+}^{2}-k_{-}^{2})}\left[ 
\begin{array}{ccc}
s^{2} & -is^{2}k^{\left( 2\right) } & 0 \\ 
is^{2}k^{\left( 2\right) } & k_{2}^{2} & 0 \\ 
0 & 0 & 0
\end{array}
\right] ,
\end{equation}
which has as eigenvalues $\lambda _{1}=0,$ $\lambda _{2}=0,$ $\lambda
_{3}=k_{2}^{2}+s^{2}$. Consequently, one has $SP>0$ (unitarity
preservation). At the pole $k_{-}^{2},$ a similar behaviour occurs: one
obtains a residue matrix exactly equal to the one given above. The
difference rests only on the coefficient appearing in front of the matrix,
in this case: $\frac{1}{D(k_{-})}%
=v_{0}^{2}[(k_{-}^{2}-s^{2})(k_{-}^{2}-k_{-}^{2})]^{-1}>0$. The fact that
this last coefficient results positive indicates that the unitarity is also
preserved at the pole $k^{2}=$ $k_{-}^{2}$, once one has the same
eigenvalues. Here, the situation presents a peculiarity with respect to its $%
\left( 1+3\right) $-dimensional counterpart: according to the analysis of
the works \cite{Adam},\cite{Baeta}, for a time-like background vector, the
gauge sector is always plagued by ghost states which cannot be removed by
any gauge choice. They actually spoil the unitarity.

\subsubsection{Space-like Case:}

In this case, taking $v^{\mu }=(0,0,v)$, the tensor $B_{\mu \nu }$ is given
as follows:

\begin{equation}
B_{\mu \nu }(k)=\frac{1}{D(k)}\left[ 
\begin{array}{ccc}
-k^{2}(\boxtimes -v^{2}k_{1}^{2}) & is\boxtimes k^{\left( 2\right)
}-k^{2}v^{2}k_{0}k^{\left( 1\right) } & ik^{\left( 1\right) }(-s\boxtimes
-sk^{2}v^{2}) \\ 
-is\boxtimes k^{\left( 2\right) }-k^{2}v^{2}k_{0}k^{\left( 1\right) } & 
k^{2}(\boxtimes +v^{2}k_{0}^{2}) & isk_{0}(\boxtimes +v^{2}k^{2}) \\ 
ik^{\left( 1\right) }(s\boxtimes +sk^{2}v^{2}) & -is\boxtimes
k_{0}+iskv^{2}k_{0} & k^{2}\left( \boxtimes +v^{2}s^{2}\right)
\end{array}
\right] ,
\end{equation}
where: $\boxtimes =k^{4}-(s^{2}-v^{2})k^{2}-v^{2}k_{2}^{2}$.

For the pole $k^{2}=0,$ with $k^{\mu }=(k_{0},0,k_{0}),$ one obtains the
same matrix attained in the time-like case, given by eq. (\ref{k2zero}).
Exactly by the same reasons presented at this former section, one can assert
that the unitarity is preserved at this pole.

For the pole $k^{2}=s^{2},$ with $k^{\mu }=(k_{0},0,k_{2})$, the resulting
matrix is identical to one given by eq. (\ref{k2s}), so that the conclusions
established in the time-like case are also valid here. The vanishing of the
saturated propagator at the pole $k^{2}=s^{2}$, in both cases, indicates
that the massive excitation $k^{2}=s^{2}$ is not dynamical in our model.

For the pole $k_{+}^{2}=$ $\frac{1}{2}\left[ \left( s^{2}+\overrightarrow{v}%
^{2}\right) \pm \sqrt{\left( s^{2}+\overrightarrow{v}^{2}\right)
^{2}+4\left( \overrightarrow{v}.\overrightarrow{k}\right) ^{2}}\right] ,$
with $k^{\mu }=(k_{0},0,k_{2}),$ the residue matrix is reduced to:

\begin{equation}
B_{\mu \nu }|_{(k^{2}=k_{+}^{2})}=\frac{v^{2}}{%
(k_{+}^{2}-s^{2})(k_{+}^{2}-k_{-}^{2})}\left[ 
\begin{array}{ccc}
0 & 0 & 0 \\ 
0 & k_{0}^{2} & isk_{0} \\ 
0 & -isk_{0} & s^{2}
\end{array}
\right] ,  \label{Kmais}
\end{equation}
where: $(k_{+}^{2}-k_{-}^{2})=\sqrt{\left( s^{2}+v^{2}\right)
^{2}+4v^{2}k_{2}^{2}}$. \ The eigenvalues of this matrix are: $\lambda
_{1}=0,$ $\lambda _{2}=0,$ $\lambda _{3}=k_{0}^{2}+s^{2}$, which leads to a
positive saturation $\left( SP>0\right) ,$ and then unitarity is guaranteed
at this pole. For the pole $k_{-}^{2},$ unitarity is also ensured, this may
be seen in an exactly similar way to the one performed in the time-like case.

Taking into account all results concerning the gauge sector of this model,
one concludes that the unitarity is preserved in both time- and space-like
cases (at all the poles of the gauge propagator) without any restriction.
Considering the restriction on the unitarity of the scalar sector at the
time-like case, one can state that our entire model preserves unitarity only
in the space-like case. It is also interesting to note that the unitarity of
the gauge sector is guaranteed even at the non-causal poles $k_{-}^{2}$,
which confirms the consistency of our model. This fact can be understood
remembering that the modes $k_{-}^{2},$ in spite of being non-causal, are
stable ones.

\section{Concluding Comments}

We have accomplished the dimensional reduction to 1+2 dimensions of a gauge
invariant, Lorentz and CPT-violating model, defined by the
Carroll-Field-Jackiw term, $\epsilon ^{\mu \nu \kappa \lambda }v_{\mu
}A_{\nu }F_{\kappa \lambda }$. One then obtains a Maxwell-Chern-Simons
planar Lagrangian in the presence of a Lorentz breaking term and a massless
scalar field. Concerning this reduced model, the CPT symmetry is conserved
for a purely space-like $v^{\mu }$, and spoiled otherwise. The propagators
of this model are evaluated and exhibit a common causal structure (bound to
the dependence on $1/\boxtimes )$. The poles of the propagators are used as
starting point for the analysis of causality, stability and unitarity.
Concerning the dispersion relations, one verifies that the modes have
positive definite energy, which ensures stability. The causality is assured
for all modes of the theory, except\ for $k_{-}^{2}$ (both in space- and
time-like case). In connection with the unitarity of this model, one has
analyzed the scalar and the gauge sectors separately, by means of the
saturation of the residue matrix. The gauge sector has revealed to be
unitary for time- and space-like background vectors, whereas the scalar
sector has showed to preserve unitarity only in the space-like case. We
should now pay attention to a special property of 3-space-time dimensions,
namely: the absence of ghosts in the gauge-field spectrum for a time-like, $%
v^{\mu }$. Unitarity is a relevant matter and an essential condition for a
consistent quantization of any theory. Once the unitarity is here ensured,
this model may become a useful and interesting tool to analyze planar
systems (including Condensed-Matter ones) with anisotropic properties.

A new version of this work \cite{Reduction2} may address the dimensional
reduction of a gauge-Higgs model \cite{Baeta} in the presence the
Carroll-Field-Jackiw term. In this case, the reduced model will be composed
by two scalar fields (one stemming from the dimensional reduction, the other
being the Higgs scalar), by a Maxwell-Chern-Simons-Proca gauge field, and by
the Lorentz-violating mixing term. The introduction of the Higgs sector may
shed light on new interesting issues concerning planar systems, like the
investigation of vortex-like configurations in the framework of a
Lorentz-breaking model.

Another natural investigation consists in studying the solutions to the
classical equations of motion (the extended Maxwell equations) and wave
equations (for the potential $A^{\mu }$) corresponding to the reduced
Lagrangian. It is possible that such equations reveal a similar structure
(but more complex) to the MCS conventional Electrodynamics, since the
reduced Lagrangian indeed contains the MCS\ sector. The solution to these
equations may unveil some interesting aspects, such as the property of
anisotropy (induced by a space-like background, $\overrightarrow{v}$) in the
interaction potential derived from such equations. This issue is actually
being investigated and we shall report on it in a forthcoming paper \cite
{Manojr}.

\section{Acknowledgments}

Manoel M. Ferreira Jr is grateful to the Centro Brasileiro de Pesquisas
F\'{i}sicas - CBPF, for the kind hospitality. \ J. A. Helay\"{e}l-Neto
acknowledges the CNPq for financial support. The authors are grateful to
\'{A}lvaro L. M. de Almeida Nogueira, Marcelo A. N. Botta Cantcheff and
Gustavo D. Barbosa for relevant discussions.

\end{document}